\documentstyle[twocolumn,floats,epsf,aps,prb]{revtex}
\begin{document}

\twocolumn[\hsize\textwidth\columnwidth\hsize\csname
@twocolumnfalse\endcsname

\title {\bf Designed nonlocal pseudopotentials for enhanced transferability}
\author{Nicholas J. Ramer and Andrew M. Rappe}
\address{Department of Chemistry and Laboratory for Research on the Structure of Matter\\ University of Pennsylvania, Philadelphia, PA 19104}
\date{\today}
\maketitle

\begin{abstract} 
A new pseudopotential generation method is presented which significantly improves transferability.  The method exploits the flexibility contained in the separable Kleinman-Bylander form of the nonlocal pseudopotential {[Phys. Rev. Lett. {\bf 48}, 1425 (1982)]}.  By adjusting the functional form of the local potential, we are able to improve the agreement with all-electron calculations.  Results are presented for the Ca atomic pseudopotential.  Configuration testing, logarithmic derivatives and chemical hardness all confirm the accuracy of these new pseudopotentials.  
\pacs{31.15.A, 71.15.Hz, 71.20.Dg}
\end{abstract} 
]
The pseudopotential (PS) approximation, or the separation of electrons into core and valence based on their level of participation in chemical bonding, lies at the heart of most modern electronic structure calculations.  The original PS formalism\cite{PhilKlein} grew out of the orthogonalized plane-wave approach.\cite{OPW}  The atomic PS replaces the nuclear Coulomb potential plus core electrons, thus simplifying the original system of differential equations.  Adopting the PS approximation may introduce some unphysical results if the PSs are not constructed judiciously.  The accuracy of the PS, or its transferability, may be gauged by its ability to reproduce the results of all-electron (AE) calculations in a variety of atomic environments.  Most transferability testing has centered on configuration testing, characterization of the scattering properties using logarithmic derivatives, and, more recently, chemical hardness.\cite{TeterPS}

The earliest PSs generated for use in density-functional theory calculations replaced the strongly attractive Coulombic potential near the origin with a weaker local potential, and core electrons were eliminated from the calculations.\cite{StarkJDJ}  In this approach, approximate agreement between PS and AE eigenvalues and logarithmic derivatives was achieved for many elements.  However, first row non-metals and first-row transition metals could not be accurately described by these PSs.

To improve PS transferability, more complicated (and more flexible) semilocal (SL) PSs were designed\cite{HSC} with a different spherically symmetric potential for each angular momentum.  This added flexibility permits the enforcement of the norm-conservation condition at the reference energy, $\varepsilon_i$, for $R$ greater than the core radius, $r_c$
\begin{eqnarray}
\left. \frac{d}{d\varepsilon}\left(\frac{d\ln\phi^{\rm{AE}}_{i}\left(\vec{r}\,\right)}{dr}\right)\right|_{R,\varepsilon_i}= \left. \frac{d}{d\varepsilon}\left(\frac{d\ln\phi^{\rm{PS}}_{i}\left(\vec{r}\,\right)}{dr}\right)\right|_{R,\varepsilon_i}
\end{eqnarray}
where $\phi^{\rm{AE}}_{i}\left(\vec{r}\,\right)$ and $\phi^{\rm{PS}}_{i}\left(\vec{r}\,\right)$ are the AE and PS Kohn-Sham eigenstates for the state $i$.  Including this criterion into PS generation greatly improves transferability.

Incorporating norm-conservation makes it possible to have exact agreement between the AE and PS eigenvalues and wave functions outside $r_c$ for one electronic reference state (RS).\cite{HSC}  However, the corresponding single-particle differential equation for a PS constructed with this method is more complicated because of the angular momentum projection.  Expressing the SL PSs within a plane-wave basis requires the computation of $V(\vec{G},\vec{G'})$ instead of just $V(\vec{G}-\vec{G'})$ where $\vec{G}$ is a reciprocal lattice vector.  This results in a huge memory expense.

The fully separable nonlocal (NL) Kleinman-Bylander PS form\cite{KB} dramatically reduces the memory cost of the SL PSs.  These PSs are constructed from a local potential and angular-momentum-dependent NL projectors.  In Fourier space, the projector can be expressed as $W(\vec{G})\cdot W(\vec{G'})$ replacing $V(\vec{G},\vec{G'})$.  This reduces the PS memory scaling from $N^2$ to $N$.  With the inclusion of the NL projectors, the resulting single-particle Kohn-Sham equation becomes an integrodifferential equation.  The solutions to an integrodifferential equation may violate the Wronskian theorem and possess noded eigenstates lower in energy than the nodeless solution.\cite{GonzeWronk}  A simple diagnostic procedure\cite{Gonzeghost} allows for detection of these lower-energy or ghost states.  The separable form of these potentials also permits efficient evaluation in solid-state calculations with $N^2$ or $N^2\log_2 N$ CPU-time scaling\cite{KSLP,Chengyu} for the NL energy contribution and its gradients.  These potentials have proven very effective for the study of computationally intensive large-scale systems.\cite{Allan}  

While configuration testing and characterization of the scattering properties are important tools in ascertaining the transferability of a PS, there are some properties that cannot be sampled using these diagnostics.  These remaining properties involve effects of electrostatic screening and non-linearity of the exchange-correlation energy.  Recently, chemical hardness conservation has been used as an effective measure of how these self-consistent terms vary with electronic configuration and as an indication of accurate PS generation.\cite{TeterPS}  The chemical hardness is defined as the matrix
\begin{eqnarray}
H_{nl,n'l'}=\frac{1}{2}\frac{\partial^2 E_{\rm{tot}}[\rho\left(\vec{r}\,\right)]}{\partial f_{nl}\partial f_{n'l'}}=\frac{1}{2}\frac{\partial\varepsilon_{nl}}{\partial f_{n'l'}}
\end{eqnarray}
where $E_{\rm{tot}}[\rho\left(\vec{r}\,\right)]$ is the total energy of the atom and is a functional of the total electronic charge density, $\rho\left(\vec{r}\,\right)$, $f_{nl}$ is the occupation number of the $nl$-th state and $\varepsilon_{nl}$ is the self-consistent $nl$-th eigenvalue.  In the second equality, we have used the fact that $\varepsilon_{nl}=\partial E_{\rm{tot}}[\rho\left(\vec{r}\,\right)]/\partial f_{nl}$.  In order to improve the chemical hardness testing results, the form of the local potential has also been studied.  Local potentials were constructed from various linear combinations of the semilocal potentials.\cite{Filippetti}  Chemical hardness testing showed that the accuracy of these PSs approached the accuracy of a SL PS from the same set of $l$-dependent potentials, but did not exceed it.

One of the other major objectives in PS generation, besides transferability, is rapid convergence in a plane-wave basis.  It has been shown that the residual kinetic energy of the RS pseudo-wavefunctions lying beyond the plane-wave cutoff energy is an excellent predictor of the basis set convergence error of the PS in a solid or molecule.\cite{RappePS}  The optimized PS construction is designed to minimize this residual kinetic energy.  

The ultra-soft PS construction\cite{VanderbiltPS} was formulated to generate highly transferable PSs with rapid convergence in a plane-wave basis.  To facilitate these improvements additional NL projectors for additional $nl$ states are included in the form of the PS.  The addition of multiple projectors per angular momentum allows exact agreement between the PS logarithmic derivatives and their slopes with the AE results for more than one energy.  In order to improve pseudo-wavefunction smoothness, the norm-conservation constraint on the wavefunction is relaxed.  To re-introduce norm conservation, a compensating valence charge density is added.

In this paper we present a method for NL PS construction according to the KB separable form which improves accuracy while retaining the convenience of a single-projector representation.  

To construct a PS, an electronic RS is chosen, and an AE calculation is performed.  Then, a PS, $\widehat{V}_{\rm{PS}}$, and pseudo-wavefunctions, $\left|\phi^{\rm{PS}}_{nl}\right>$, are chosen which satisfy 
\begin{eqnarray}
\left(\widehat{T}+\widehat{V}_{\rm{H}}[\rho]+\widehat{V}_{\rm{XC}}[\rho]+\widehat{V}_{\rm{PS}}\right)\left|\phi^{\rm{PS}}_{nl}\right>=\varepsilon_{nl}\left|\phi^{\rm{PS}}_{nl}\right>
\end{eqnarray}
\noindent where $\widehat{T}$ is the single-particle kinetic energy operator and $\widehat{V}_{\rm{H}}[\rho]$ and $\widehat{V}_{\rm{XC}}[\rho]$ are the self-consistent Hartree and exchange-correlation energy operators, respectively.  The latter two operators are functionals of the total charge density, $\rho\left(\vec{r}\,\right)$, where $\rho\left(\vec{r}\,\right)=\sum_{nl}f_{nl}\left|\phi^{\rm{PS}}_{nl}\left(\vec{r}\,\right)\right|^2$.  We require that the pseudo-wavefunctions obey the following criteria:
\begin{enumerate}
\item \noindent $\phi^{\rm{PS}}_{nl}(\vec{r}\,)=\phi^{\rm{AE}}_{nl}(\vec{r}\,)$ for $r>r_c$
\item $\varepsilon^{\rm{PS}}_{nl}=\varepsilon^{\rm{AE}}_{nl}$
\item $\left. \frac{d}{d\varepsilon}\left(\frac{d\ln\phi^{\rm{AE}}_{nl}(\vec{r}\,)}{dr}\right)\right|_{R,\varepsilon_{nl}}= \left. \frac{d}{d\varepsilon}\left(\frac{d\ln\phi^{\rm{PS}}_{nl}(\vec{r}\,)}{dr}\right)\right|_{R,\varepsilon_{nl}}$
\item $\left<\phi^{\rm{PS}}_{nl}\right|\left.\phi^{\rm{PS}}_{nl}\right>=\left<\phi^{\rm{AE}}_{nl}\right|\left.\phi^{\rm{AE}}_{nl}\right>=1$.
\end{enumerate}
If $\widehat{V}_{\rm{PS}}$ is not $l$-dependent, the resulting PS is called local (radially and angularly local).  More generally, $\widehat{V}_{\rm{PS}}$ can be separated into a local potential and a sum of short-ranged corrections:
\begin{eqnarray}
\widehat{V}_{\rm{PS}}=\widehat{V}^{\rm{loc}}+\sum_{l}\Delta\widehat{V}_{l}
\end{eqnarray}
\noindent where
\begin{eqnarray}
\widehat{V}^{\rm{loc}}\equiv\int d^{3}r\left|\vec{r}\,\right>V^{\rm{loc}}\left(\vec{r}\,\right)\left<\vec{r}\,\right|.
\end{eqnarray}
\noindent For a SL PS, the corrections, $\Delta\widehat{V}_{l}^{\rm{SL}}$, are projection operators in the angular coordinates and local in the radial coordinate.  To construct a fully separable NL PS, $\Delta\widehat{V}_{l}$ is formed according to:
\begin{eqnarray}
\Delta \widehat{V}^{\rm{NL}}_{l}\equiv\frac{\Delta\widehat{V}_{l}^{\rm{SL}}\left|\phi^{\rm{PS}}_{nl}\right>\left<\phi^{\rm{PS}}_{nl}\right|\Delta\widehat{V}_{l}^{\rm{SL}}}{\left<\phi^{\rm{PS}}_{nl}\left|\Delta\widehat{V}_{l}^{\rm{SL}}\right|\phi^{\rm{PS}}_{nl}\right>}.
\end{eqnarray}
\noindent When $\widehat{V}_{\rm{PS}}$ operates on $\left|\phi^{\rm{PS}}_{nl}\right>$ we obtain
\begin{eqnarray}
\widehat{V}_{\rm{PS}}\left|\phi^{\rm{PS}}_{nl}\right>=\left(\widehat{V}^{\rm{loc}}+\Delta\widehat{V}_{l}^{\rm{SL}}\right)\left|\phi^{\rm{PS}}_{nl}\right>.
\end{eqnarray}
Therefore for the RS we are guaranteed exact agreement between the eigenvalues and wavefunctions (beyond $r_c$) of the SL and NL atoms. 

\begin{table*}[t]
\caption{Configuration testing for the Ca atom.  Eigenvalues and $\Delta E_{\rm{tot}}$ are given for an all-electron (AE) potential.  The reference state is 3$s^2$3$p^6$4$s^0$3$d^0$.  Absolute errors are given for a semilocal (SL) pseudopotential, an undesigned nonlocal (UNL) pseudopotential and a designed nonlocal (DNL) pseudopotential generated with the presented method.  All energies are in Ry.}
\begin{tabular}{cdddd|cdddd}
State&AE&SL&UNL&DNL&State&AE&SL&UNL&DNL\\ 
 &Energy&Error&Error&Error& &Energy&Error&Error&Error \\ \hline  
 & & & & & & & & & \\
               $3s^2$&-4.5277& 0.0000& 0.0000& 0.0000&               $3s^2$&-3.2284& 0.0022& 0.0021& 0.0005\\  
               $3p^6$&-3.1688& 0.0000& 0.0000& 0.0000&               $3p^6$&-1.8875& 0.0024& 0.0023& 0.0007\\  
               $4s^0$&-1.0537&-0.0099&-0.0099& 0.0000&               $4s^1$&-0.2469&-0.0023&-0.0022& 0.0001\\  
               $3d^0$&-1.1933& 0.0000& 0.0000& 0.0000&               $3d^1$&-0.0648& 0.0012& 0.0012& 0.0000\\ 
$\Delta E_{\rm{tot}}$& 0.0000& 0.0000& 0.0000& 0.0000&$\Delta E_{\rm{tot}}$&-1.1903&-0.0035&-0.0036&-0.0001\\  
 & & & & & & & & & \\
               $3s^2$&-3.9220& 0.0059& 0.0059& 0.0000&               $3s^2$&-4.4495& 0.0198& 0.0199& 0.0003\\  
               $3p^6$&-2.5681& 0.0057& 0.0057& 0.0000&               $3p^5$&-3.0670& 0.0191& 0.0192& 0.0000\\  
               $4s^1$&-0.6716&-0.0039&-0.0039& 0.0000&               $4s^2$&-0.8070&-0.0045&-0.0045& 0.0007\\  
               $3d^0$&-0.6401& 0.0044& 0.0044& 0.0000&               $3d^0$&-1.0294& 0.0157& 0.0156& 0.0006\\ 
$\Delta E_{\rm{tot}}$&-0.8746&-0.0062&-0.0062& 0.0000&$\Delta E_{\rm{tot}}$& 1.2031&-0.0223&-0.0223&-0.0003\\  
 & & & & & & & & & \\
               $3s^2$&-3.4115& 0.0093& 0.0093& 0.0000&               $3s^2$&-5.0789& 0.0138& 0.0139& 0.0004\\  
               $3p^6$&-2.0601& 0.0089& 0.0089& 0.0000&               $3p^5$&-3.6924& 0.0134& 0.0135& 0.0009\\  
               $4s^2$&-0.2833&-0.0011&-0.0011& 0.0000&               $4s^1$&-1.2845&-0.0095&-0.0095& 0.0001\\   
               $3d^0$&-0.1659& 0.0064& 0.0064& 0.0000&               $3d^0$&-1.6335& 0.0112& 0.0111& 0.0009\\ 
$\Delta E_{\rm{tot}}$&-1.3478&-0.0086&-0.0086& 0.0000&$\Delta E_{\rm{tot}}$& 2.2464&-0.0154&-0.0155&-0.0003\\  
\end{tabular}
\end{table*}

However, for any state, $\left|\psi^{\rm{PS}}_{n'l}\right>$, other than the RS:
\begin{eqnarray}
\widehat{V}_{\rm{PS}}\left|\psi^{\rm{PS}}_{n'l}\right>\neq\left(\widehat{V}^{\rm{loc}}+\Delta\widehat{V}_{l}^{\rm{SL}}\right)\left|\psi^{\rm{PS}}_{n'l}\right>
\end{eqnarray}
\noindent where $n'$ is not required to equal $n$.  The inequality in equation (8) illustrates the difficulties involved in assessing and improving the transferability of NL PSs:  the transferability of a NL PS can be dramatically different from the corresponding SL PS.  

For simplicity, we have focused on the single-projector NL PS construction in the current approach.  We believe that implementation of the presented method along with the ultra-soft construction and multiple projectors, may provide even greater transferability.  We begin by constructing an optimized SL PS.  Since equation (4) is simply an addition of local and NL terms, we may alter $\widehat{V}^{\rm{loc}}$ arbitrarily without losing the exact agreement between the AE and NL eigenvalues and pseudo-wavefunctions at the RS, $provided$ we adjust the NL corrections accordingly.  However, we change the eigenvalue agreement at any other configuration by doing so.  

Operationally, an additional electronic configuration or design state (DS) is chosen.  A local augmentation operator ($\widehat{A}$) is added to the local potential forming a designed nonlocal (DNL) PS.  The augmentation operator is subtracted from the NL corrections $\Delta\widehat{V}_{l}$ in the following way:
\begin{eqnarray}
\widehat{V}_{\rm{PS}}=\left(\widehat{V}^{\rm{loc}}+\widehat{A}\right)+\sum_l \Delta\widehat{V}^{\rm{DNL}}_{l}
\end{eqnarray}
\noindent where
\begin{eqnarray}
\Delta\widehat{V}^{\rm{DNL}}_{l}=\frac{\left(\Delta\widehat{V}_{l}^{\rm{SL}}-\widehat{A}\right)\left|\phi^{\rm{PS}}_{nl}\right>\left<\phi^{\rm{PS}}_{nl}\right|\left(\Delta\widehat{V}_{l}^{\rm{SL}}-\widehat{A}\right)}{\left<\phi^{\rm{PS}}_{nl}\right|\left(\Delta\widehat{V}_{l}^{\rm{SL}}-\widehat{A}\right)\left|\phi^{\rm{PS}}_{nl}\right>}. 
\end{eqnarray}
When the augmented $\widehat{V}_{\rm{PS}}$ operates on the RS, the result reduces to equation (7).  However for any state other than the RS, the second term in equation (9) will contribute differently.

By adjusting $\widehat{A}$, it is possible to obtain almost exact agreement between the AE and DNL eigenvalues for a particular DS.  With the proper selection of DS electronic configuration, excellent transferability can be obtained for a variety of ionized and excited states.

For the Ca atom, we have chosen a RS which is highly ionized and a DS which is neutral.  All atomic energy calculations were done within the LDA and optimized PS generation methods were used.  A 50 Ry plane-wave cutoff energy was used to achieve excellent plane-wave convergence of the pseudo-wavefunctions for the atom.  We have chosen the $s$ angular momentum channel to be the foundation for the local potential and a square step potential barrier as our augmentation operator.  The height and radius of the step have been adjusted to reproduce the AE eigenvalues for the DS.  

Our selection of RS deserves additional mention.  For Ca, semi-core states were included as valence in the PS.  The inclusion of these states allowed for the removal of ghost levels, greater local potential design flexibility and better overall transferability of the PS.  It is also important to note that although we have included multiple $s$-channel states, we do $not$ treat these states with different projection operators.  The PSs are generated using only one NL projector for each angular momentum channel.

\begin{table*}[t]
\caption{Chemical hardness testing for the Ca atom.  Absolute hardness values are compared for an all-electron (AE), semilocal pseudopotential (SL), an undesigned nonlocal (UNL) pseudopotential and a designed nonlocal (DNL) pseudopotential generated with the presented method.  Hardness values were determined for two different electronic configurations.  Each element of the symmetric hardness matrix, $H_{nl,n'l'}$, is indexed by the change in the $nl$-th eigenvalue for a change of the $n'l'$-th occupation number.  Units are in Ry per occupation number.}
\begin{tabular}{rr|cccc|cccc}
 & & &\multicolumn{2}{c}{3$s^{1.95}$3$p^{5.9}$4$s^1$3$d^{0.1}$}& & &\multicolumn{2}{c}{3$s^{2}$3$p^{6}$4$s^{2}$3$d^{0.01}$}&\\ 
$nl$&$n'l'$&AE&SL&UNL&DNL&AE&SL&UNL&DNL\\ \hline
3$s$&3$s$&0.5655&0.5621&0.5620&0.5657&  0.4945&0.4897&0.4897&0.4946\\  
    &3$p$&0.5474&0.5442&0.5441&0.5475&  0.4767&0.4723&0.4722&0.4767\\  
    &4$s$&0.2830&0.2853&0.2853&0.2830&  0.2257&0.2269&0.2269&0.2257\\ 
    &3$d$&0.4614&0.4587&0.4586&0.4614&  0.3677&0.3635&0.3634&0.3677\\  

3$p$&3$p$&0.5310&0.5279&0.5279&0.5309&  0.4607&0.4566&0.4565&0.4606\\  
    &4$s$&0.2813&0.2835&0.2835&0.2813&  0.2250&0.2262&0.2262&0.2250\\ 
    &3$d$&0.4506&0.4482&0.4482&0.4507&  0.3593&0.3555&0.3554&0.3594\\  

4$s$&4$s$&0.2079&0.2097&0.2097&0.2079&  0.1790&0.1800&0.1800&0.1790\\ 
    &3$d$&0.2639&0.2655&0.2655&0.2639&  0.2101&0.2106&0.2106&0.2101\\  

3$d$&3$d$&0.3941&0.3916&0.3915&0.3936&  0.2989&0.2952&0.2951&0.2986\\
\end{tabular}
\end{table*}

The RS for the Ca PS is 3$s^2$3$p^6$4$s^0$3$d^0$. The core regions extend to 1.29, 1.60 and 1.27 a.u. for the $s$, $p$ and $d$ states, respectively.  We selected a neutral electronic configuration (3$s^2$3$p^6$4$s^2$3$d^0$) as our DS.  We found that a square potential step with height of 6.76 Ry and radius of 0.93 a.u. as the augmentation operator gives excellent transferability.

Table I contains configurational testing results for the Ca atom.  We present results from an AE atom, a SL PS, an undesigned NL (UNL) PS generated with an unaugmented local potential consisting of only the $s$ angular momentum channel potential, and a DNL PS which includes the square step.  The potentials were tested in both ionized and excited electronic configurations.  In addition to eigenvalue information, the table also compares total energy differences from the RS.  For our choice of augmentation operator, we find excellent agreement in eigenvalues and total energy differences.  In all configurations shown, the errors are below 1 mRy for the DNL results.  The eigenvalue and total energy errors for the DNL potential are one to two orders of magnitude smaller than the SL and UNL results.  

Figure 1 shows the logarithmic derivative differences between the AE potential and all three PS generation methods for the $s$, $p$ and $d$ potentials of the Ca atom.  For the $s$ and $p$ channels, we find excellent agreement over a large energy range between the PSs generated with the DNL method and the AE potentials.  The energy errors in the $s$ and $p$ states for the SL and UNL PSs presented in Table I are directly related to the logarithmic derivative differences.  Interestingly, we find that the DNL PS $d$ state logarithmic derivative differs from the AE results more than the other methods for the RS.  This finding suggests that hardness testing may be a better predictor of transferability than logarithmic derivative testing, and we plan to investigate this point further.  In analogous logarithmic derivative testing at other electronic configurations, the DNL PS $d$ state scattering properties are more accurate than the other PSs in the range of energies shown in Table I.

Table II contains chemical hardness testing results for Ca AE and pseudoatoms, for two electronic configurations.  In both cases, we find excellent agreement in chemical hardness between the AE and DNL potentials for each electronic configuration.  With errors below 0.5 mRy, the DNL PS is one to two orders of magnitude more accurate than the SL and UNL PSs.

In this paper, we have developed and implemented a fully nonlocal pseudopotential approach using the separable form of Kleinman and Bylander.  We have exploited the implicit flexibility contained within the separation of the pseudopotential into local and nonlocal parts by including an augmentation operator into the local and nonlocal parts of the potential.  By adjusting the augmentation operator, we have been able to achieve almost exact agreement with all-electron results for a variety of ionized and excited states.  In addition to configuration testing, we have presented logarithmic derivatives and chemical hardness tests.  All the tests demonstrate significant improvement over semi-local and standard nonlocal potentials.  Furthermore, we have shown that for electronic configurations that contain multiple states with the same angular momentum, it is possible to construct a pseudopotential with a single nonlocal projector that will yield very accurate results.  
      
The authors are thankful for helpful discussions with S. P. Lewis during the preparation of this manuscript.  This work was supported by the Laboratory for Research on the Structure of Matter and the Research Foundation at the University of Pennsylvania as well as NSF grant DMR 97-02514 and the Petroleum Research Fund of the American Chemical Society (Grant No. 32007-G5).  Computational support was provided by the San Diego Supercomputer Center.

\begin{figure*}[b]
\epsfxsize=6.5in
\centerline{\epsfbox[18 144 592 718]{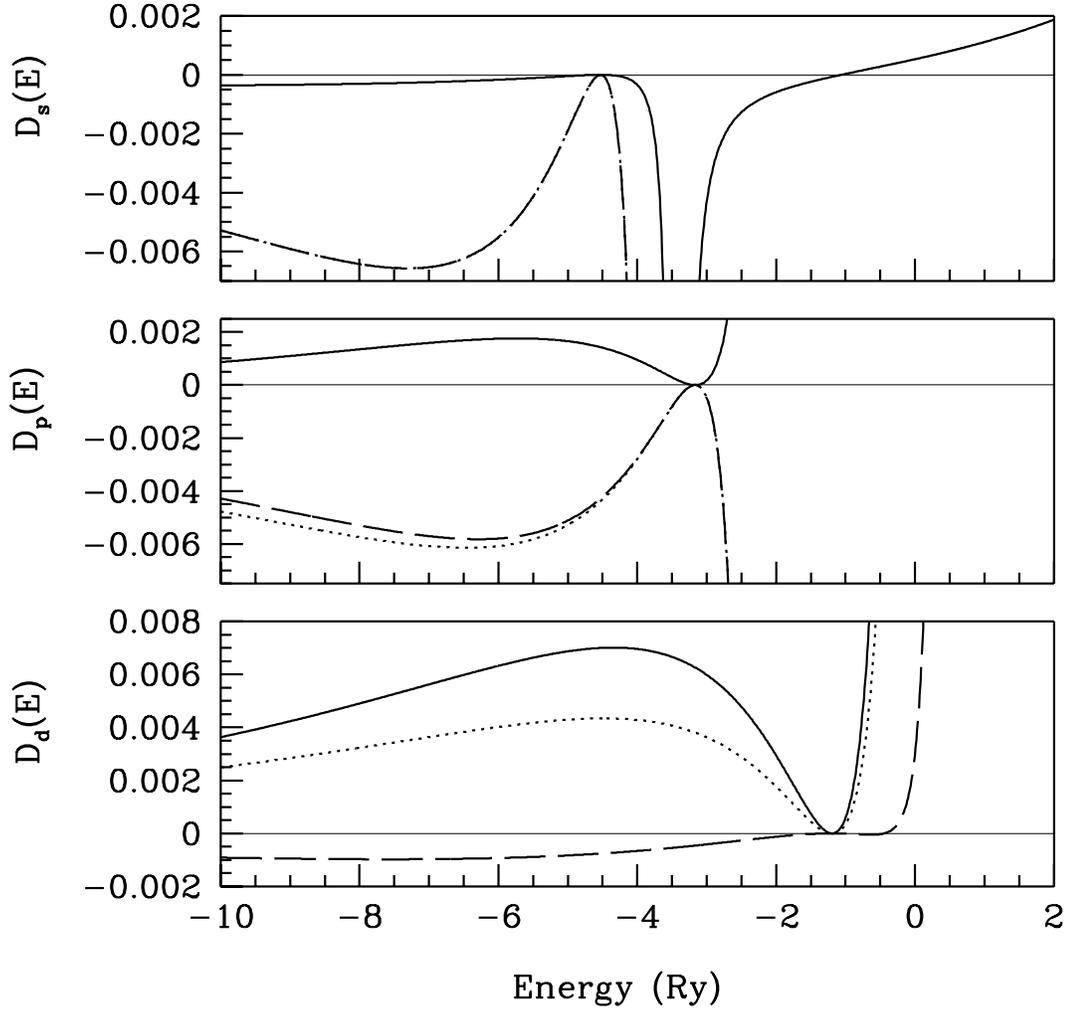}}
\caption{
Differences in logarithmic derivatives between the Ca pseudopotential and the all-electron calculation for the $s$ (top), $p$ (middle) and $d$ (bottom) states at the reference configuration (3$s^2$3$p^6$4$s^0$3$d^0$).  Results are given for a semilocal pseudopotential (dotted line), an undesigned nonlocal pseudopotential (dashed line) and a designed nonlocal pseudopotential generated with the presented method (solid line).}
\end{figure*}

\newpage
\end{document}